\documentclass[referee]{raa}
\usepackage{graphicx,times}
\usepackage{amssymb,amsmath}
\usepackage{color}

\begin{document}

\title{Kinematics of Solar neighborhood stars and its dependency on age and metallicity}

 \volnopage{ {\bf 2014} Vol.\ {\bf X} No. {\bf XX}, 000--000}
   \setcounter{page}{1}

   \author{Zong-Bo Huyan\inst{1,2}, Zi Zhu\inst{1,2}, Jia-Cheng Liu\inst{1,2}}

  \institute{ School of Astronomy and Space Science, Nanjing University, 210093 Nanjing,
China; {\it huyanzongbo@hotmail.com}\\
   \and
   Key Laboratory of Modern Astronomy and Astrophysics, Ministry of Education
    (Nanjing University), 210093 Nanjing, China\\
}
   \date{Received~~2014 month day; accepted~~2014~~month day}

\abstract{We have constructed a catalog containing best available astrometric, photometric, radial velocity and astrophysical data for mainly F-type and G-type stars (called the Astrometric catalog associated with Astrophysical Data, ACAD), which contains 27,553 records, and is used for the purpose of analyzing the stellar kinematics in the Solar neighborhood. Using the Lindblad-Oort Model and compiled ACAD, we calculated the Solar motion and Oort constants in different age/metallicity bins. The evolution of kinematical parameters with stellar age and metallicity were investigated directly. The results show that the component of the Solar motion in the direction of Galactic rotation (denoted $S_2$) has a linear increase with respect to age, which may be a consequence of the scattering processes, and its value for a dynamical cold disk was found to be $8.0\pm1.2~\mathrm{km~s^{-1}}$. $S_2$ also increases linearly with respect to metallicity, which indicates that radial migration is correlated to the metallicity gradient. On the other hand, the rotational velocity of the Sun around the Galactic center has no clear correlation with ages or metallicities of stars used in the estimation.
}

   \authorrunning{Z.-B. Huyan et al.}            
   \titlerunning{Age \& metallicity dependent kinematics}  
   \maketitle

\section{Introduction}
\label{sec:intro}

Positions, velocities, ages and chemical compositions of stars in the Solar neighborhood are fundamental data for testing the models of local Galactic kinematics and evolution. Therefore a comprehensive catalog with all these stellar parameters are necessary for the purpose of studying our Galaxy.

Since the Hipparcos catalog has provided high-precision three-dimensional positions and proper motions for nearby stars, radial velocities turn out to be important supplementary material for estimating more accurately the Galactic kinematics.In the last ten years, some new radial velocity data were obtained and some old radial velocity data were improved. Gontcharov (\cite{PCRV}) built The Pulkovo Compilation of Radial Velocities (PCRV) which contains weighted mean radial velocities for 35,495 Hipparcos stars. The Radial Velocity Experiment (RAVE), a milestone survey aiming at measuring the radial velocity of stars in the thick disk, was constructed by Siebert et al.~(\cite{RAVE}) and the catalog has been updated to its third version (Siebert et al.~\cite{RAVEIII}). Ages and metallicities, regarded as evolution index of stars, are also important. Several catalogs with astrophysical data have been presented since 1990s. Considering that a number of kinematical analysis for K, M giants have been operated (e.g. Miyamoto et al.~\cite{Miyamoto93}, Famaey et al.~\cite{Famaey05}), we focused on collecting F-type and G-type stars in the Solar neighborhood.

A very important work on kinematical evolution by Dehnen \& Binney (\cite{Binney98}) has studied the relationship between stellar kinematics and velocity dispersions/colors based on the Hipparcos proper motions. It shows that $S_{2}$ (the Solar motion along the direction of the Galactic rotation) is related to the ages of the stars which form the Local Standard of Rest (LSR). We mention that no radial velocity data were used in that work and the method of estimation is more or less indirect.

In this paper, we collect currently best astrometric data, radial velocities and astrophysical data of F-type and G-type stars to construct a comprehensive catalog, called the Astrometric catalog associated with Astrophysical Data (ACAD). In the second step, the Galactic kinematics are studied with the Lindblad-Oort model using various stellar parameters in ACAD. Then we obtain the relations between kinematics and age/metallicity and compared our result with previous works.

\section{Catalog compilation}
\label{sec:composition}

We compiled our catalog (ACAD) based on of the following three main catalogs for F and G type stars:
\begin{itemize}
\item The Geneva-Copenhagen Survey of the Solar neighborhood (GCS, Nordstr\"{o}m et al.~\cite{GCS}). GCS aimed at consolidating the calibrations of uvby photometry into distance, metallicity, and age for F-type and G-type stars. Radial velocities in GCS are provided mainly by using the CORAVEL observations and were standardized to the IAU or Wilson velocity system.
\item Marsakov and Shevelev~(\cite{v89}) Catalog of ages, metallicities, orbital elements, and other parameters for nearby F type stars (denoted MS95). MS95 contains metallicities for all the 5,498 stars and ages for 3,405 slightly evolved stars. However the space velocities relative to the Sun are only available for 1,787 stars.
\item The Ammons S. M. et al.~(\cite{v136}) N2K Consortium. IV. New temperatures and metallicities for more than 100,000 FGK dwarfs (denoted AM06). Ammons et al.~(\cite{v136}) used an uniform procedure to estimate fundamental stellar properties of Tycho-2 stars (H{\o}g et al.~\cite{Tycho-2}). However, radial velocity and age data were not included in AM06.
\end{itemize}

By setting 0.1 arc second as the upper limit of the separation of the spherical position for common star, we cross-identified stars in the above three catalogs. We found that MS95 has no intersection with the other two catalogs, while AM06 and GCS have fewer than 100 stars in common. For these stars, we used the GCS data because no age data are available in AM06 and the age/metallicity data in GCS are more accurate. The combined catalog (by cross-identification) of these three individual catalogs are the basis of our catalog.

However, this preliminary combined catalog still need revisions and supplements to reach the final catalog for the following reasons. Firstly, since the distances of stars in the three original catalogs are derived from photometric data, the results depend on different models and measurements of colors. Therefore distances from three catalogs are not consistent: they need to be revised in order to improve the accuracy and reliability. Secondly, radial velocities for almost half of the stars are not available from the three catalogs. In addition, updates of the current radial velocities are necessary for better accuracy. Finally, photometric data should be supplemented to determine the probable position of stars on the Hertzsprung-Russel diagram. In order to accomplish all these revisions and supplements, the same standard (0.1 arcsec match radius criterion) was used whenever cross-identification is used to find common  stars in different catalogs.

\subsection{Improve the distance data}

The distances of the stars in MS95 are photometric distances derived from absolute magnitudes based on Crawford (\cite{Crawford}) and Olsen (\cite{Olsen}) and visual magnitudes. The absolute magnitudes and distances in GCS were obtained with a similar method of MS95. The uncertainty of photometric distance in GCS is around 13\%. AM06 is a catalog built on Tycho-2 (H{\o}g et al.~\cite{Tycho-2}) system. Ammons et al.~(\cite{v136}) estimated the distances for all the stars in Tycho-2 with the relation fitted from parallaxes, colors and proper motions in the Hipparcos catalog and the average error of the distances in AM06 is about 100\%.

Since the accuracy of the Hipparcos parallaxes is better than most of the distances data in the three basic catalogs, the parallaxes from Hipparcos catalog can be used to improve the accuracy. The distance data in the combined catalog were replaced with Hipparcos measurements if (1) The error of Hipparcos parallax is smaller than 25\%, and (2) The error is smaller than 3 $mas$. Note that the distances in the three basic catalogs are either fitted from the Hipparcos catalog or have no color-dependent bias with Hipparcos parallaxes (Nordstr\"om et al. \cite{GCS}), all the distances are consistent with the Hipparcos system.

\subsection{Supplementary radial velocities}
Including three-dimensional velocities is one of the important features of our catalog ACAD. For radial velocity data, we should make sure that there exist no systematic difference between various input data. In our compilation, the following three catalogs were selected for the supplement of the radial velocities:
\begin{itemize}
\item The General Catalog of Radial Velocities (GCRV, Barbier-Brossat \& Figon~\cite{GCRV}). The radial velocities are derived from spectrum and standardized to the IAU or Wilson velocity systems.
\item Pulkovo radial velocities for 35,493 HIP stars (PCRV, Gontcharov~\cite{PCRV}). PCRV is a combination of 203 publications. 12 are major catalogs which provid the most (80\%) part of data in PCRV. The GCRV is one the 12 major catalogs: some stars in GCRV are used as a standard for adjusting the radial velocities in PCRV in the framework of IAU standard.
\item The third release of Radial Velocity Experiment (RAVEIII, Siebert et al.~\cite{RAVEIII}), which is focus on the area away from the plane of the Milky Way ($|b| > 25^{\circ}$) and on stars with apparent magnitudes $9 < I_{\rm DENIS} < 13$. The mean error of its radial velocities is $\sim 2\rm{km~s^{-1}}$.
\end{itemize}

All the radial velocities provided by the above catalogs are adjusted to the IAU standard system with high precision, therefore are homogeneous and self-consistent. For common stars from cross-identification, the source of radial velocity data was chosen by the priority order of GCRV, PCRV, RAVEIII. All the possibilities for space velocity are shown in Table~\ref{tbl:RVstatistic}. The positions and proper motions for all the stars from various catalogs are in the Hipparocs or Tycho-2 system, which are also known as the International Celestial Reference Frame (ICRF) in optical bandpass.  We can conclude that the 6 dimensional space velocity data (including five astrometric parameters and one radial velocity) in the final catalog are systematically homogeneous.

\begin{table}
\begin{center}
\caption{Statistics of space velocity data in ACAD}
\label{tbl:RVstatistic}
 \begin{tabular}{cllr}
  \hline\noalign{\smallskip}
Index & Proper motion  & Radial velocity & Number \\
  \hline\noalign{\smallskip}
1  &  HIP  &  PCRV  &  233\\
2  &  HIP  &  GCS   &  10453\\
3  &  HIP  &  GCRV  &  554\\
4  &  MS95 &  MS95  &  1787\\
5  &  HIP  & RAVE  &  40\\
6  &  TYCHO-2  & RAVE & 14486\\
  \noalign{\smallskip}\hline
\end{tabular}
\end{center}
\end{table}

\subsection{Photometric data in ACAD}
The absolute magnitude and spectral type were also provided in ACAD. For the stars recognized as the Hipparcos stars, we derived the absolute magnitude from the Hipparcos parallax and Hp magnitude (visual magnitude in Hipparcos system). Among the rest, the the original absolute magnitudes from MS95 and GCS catalog of stars are used, while for AM06 stars, Tycho V magnitude are used in computing the absolute magnitude with distance data.

The collected spectral type in ACAD was described by MK classification. The spectral type is composed of two parts: temperature class as an indication of the surface temperature and luminosity class as an indication of surface gravity. For the stars without spectral type in the combined catalog, the Tycho-2 Spectral Type catalog (SpType) by Wright et al. (\cite{SpType}) was used as a supplement.

\subsection{Age and metallicity}
GCS has been updated twice to GCSIII (Holmberg et al.~\cite{GCSIII}) and metallicities were re-calibrated by using the infrared flux method (Casagrade et al.~\cite{Casagrande}). However, we prefer to use GCS in this version for the following two reasons: (1) GCS are consistent with several previous works (e.g. Edvardsson et al.~\cite{Edvard}) and proved to be reliable enough. The GSCIII catalog need to be further re-checked with independent methods for calculating age and metallicity before applied as a part of ACAD: this will be done in the near future for the next version of ACAD. (2) The differences of age/metallicity between GCS and GCSIII are not quite significant, the largest change in ages is only $\sim 10\%$. For most analysis of the kinematical evolution, this would not affect the results systematically.

\subsection{Statistics of ACAD}
\begin{figure}
\centering
\includegraphics[width=\textwidth, angle=0]{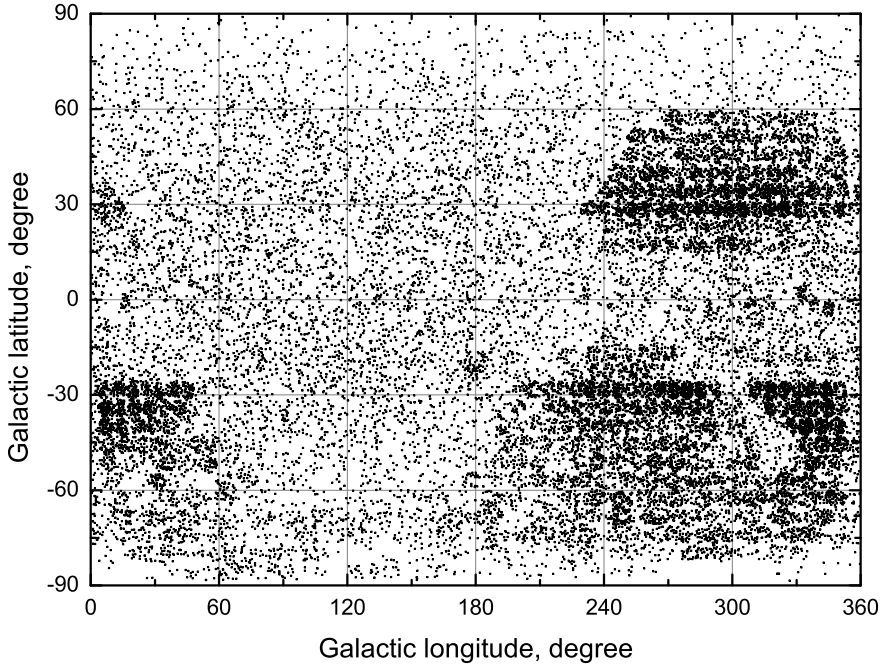}
\caption{The distribution of the 27,553 ACAD stars in the Galactic coordinate system.}
\label{pic:distribution}
\end{figure}

ACAD is a combination of 8 different catalogs (3 basic catalogs for F-type and G-type stars, the Hipparcos catalog for the revision of distances, 3 radial velocity catalogs for supplements and revisions of space velocity and 1 catalog for the supplement of spectral type). All the stars are filled with distance, space velocity and at least one type of astrophysical data (age and metallicity). For all the spectra, about $23\%$ are G-type, $75\%$ are F-type and the rest are K-type or A-type. The distribution of the ACAD stars is plotted in Figure~\ref{pic:distribution}. The high-density region with a latitude from $-60\deg$ to $-20\deg$ and from $20\deg$ to $60\deg$ are covered by RAVE which targeted on the thick disk. There are 27,553 stars in total and the statistics of different parameters are listed in Table~\ref{tbl:statistic}.

\begin{table}
\begin{center}
\caption{Statistics of different parameters of stars in ACAD}
\label{tbl:statistic}
 \begin{tabular}{ll}
  \hline\noalign{\smallskip}
Description of data      & Number               \\
  \hline\noalign{\smallskip}
Total number of ACAD stars        &  27,553  \\
stars with absolute magnitude   &  27,553     \\
stars with spectral type     &    9,443      \\
stars with space velocities   &  27,553     \\
stars with distances\&space velocities\&ages     & 12,464    \\
stars with distances\&space velocities\&[Fe/H]   & 27,553    \\
  \noalign{\smallskip}\hline
\end{tabular}
\end{center}
\end{table}

\begin{figure}
\centering
\includegraphics[width=\textwidth, angle=0]{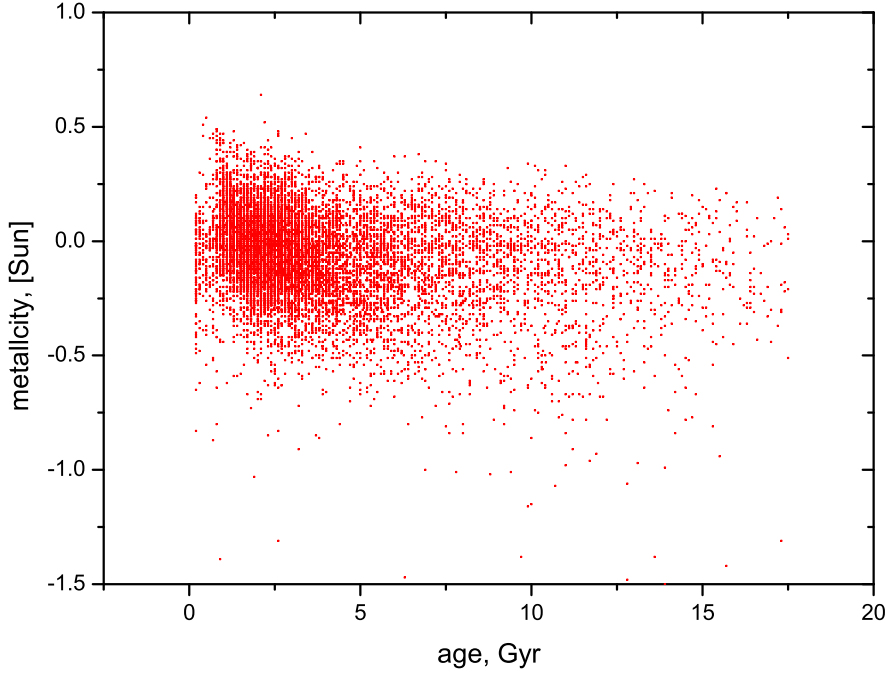}
\caption{Metallicity versus age for ACAD stars.}
\label{pic:Fe_Age}
\end{figure}

The distances of all the stars in ACAD range from 1.3 pc to 446.4 pc and $84.6\%$ are within 100 pc. The youngest star in ACAD is 0.2 Gyr old and the oldest is 17.0 Gyr. Over $60\%$ of ACAD stars are younger than the Sun ($4.6\pm 0.1\mathrm{ Gyr}$, Bonanno~\cite{Bonanno}) and almost half of the stars are between 1-3 Gyr old. The metallicities range from -3.2 [Sun] to 3.5 [Sun] and centralize at 0 [Sun] (the metallicity of the Sun). Figure~\ref{pic:Fe_Age} is the correlation of age and metallicity in ACAD, which shows that the two parameters are not obviously related. In the following sections, we will study the kinematical properties of the stars in the Solar neighborhood by using a sample selected from ACAD.

\section{Kinematical models}
\label{sec:models}
To describe the characteristics of stellar motion in our Galaxy, it is practical to introduce the Galactocentric cylindrical coordinate system ($R,~\theta,~\mathit{z}$) where $R$ is the radius from Galactic center, $\theta$ the azimuthal angle measured from the axis directing to the Galactic center from the Sun, and $z$ the distance to the Galactic plane. Accordingly, we have $V_R$ (the radial velocity with respect to the Galactic center), $V_{\theta}$ (the rotation projected to the Galactic plane) and $V_z$ (the velocity perpendicular to the Galactic plane). However, neither the velocity $\mathbf{V}(V_R, V_{\theta}, V_z)$ nor the position $(R, \theta, z)$ can be obtained directly from the ACAD catalog because all the measurements are operated with respect to the Sun. It is necessary to establish a connection between the Galactocentric cylindrical coordinate system and the heliocentric Galactic coordinate system. Velocity field ($V_R, V_{\theta}, V_z$) can be expanded in the Galactocentric cylindrical coordinate system as follows:
\begin{equation}
\begin{pmatrix}
V_R \\
V_{\theta} \\
V_{z} \\
\end{pmatrix}
=
\begin{pmatrix}
V_{R_{\bigodot}} \\
V_{\theta_{\bigodot}} \\
V_{z_{\bigodot}} \\
\end{pmatrix}
-
\begin{pmatrix}
S_1 \\
S_2 \\
S_3 \\
\end{pmatrix}
+
\mathbf{M}
\begin{pmatrix}
\delta R \\
\delta \theta \\
\delta z \\
\end{pmatrix},
\label{eq:tylor}
\end{equation}
where $\mathbf{V_{\bigodot}} = (V_{R_{\bigodot}}, V_{\theta_{\bigodot}}, V_{z_{\bigodot}})^{T}$ is the Solar velocity in the Galactocentric cylindrical coordinate system; $\mathbf{S} = (S_{1}, S_{2}, S_{3})$ is the Solar motion with respect to Local Standard of Rest (LSR). The matrix $\mathbf{M}$ is such that
\begin{equation}
\mathbf{M} = \mathbf{M}^{+} + \mathbf{M}^{-},
\label{eq:M+M-}
\end{equation}
where $\mathbf{M}^{-}$ is a matrix representing the average angular velocity of all the stars as a rigid-body. $\Omega_{3}$ (equals to Oort Constant $A$) is the angular velocity around the Galactic pole ($z$ direction), which is more significant than the other two components perpendicular to $z$ axis (Zhu~\cite{Zhu00}, Miyamoto et al.~\cite{Miyamoto93}). $\mathbf{M}^{-}$ can be written as:
\begin{equation}
\mathbf{M}^{-}
=
\begin{pmatrix}
0 & -\Omega_{3} & \Omega_{2} \\
\Omega_{3} & 0 & -\Omega_{1} \\
-\Omega_{2} & \Omega_{1} & 0 \\
\end{pmatrix}.
\label{eq:M-}
\end{equation}
$\mathbf{M}^{+}$ is a symmetric matrix characterizing the deformation (oblique shear and dilatation) of the velocity field:
\begin{equation}
\mathbf{M}^{+}
=
\begin{pmatrix}
M_{11}^{+} & M_{12}^{+} & M_{13}^{+} \\
M_{12}^{+} & M_{22}^{+} & M_{32}^{+} \\
M_{13}^{+} & M_{32}^{+} & M_{33}^{+} \\
\end{pmatrix}.
\label{eq:M+}
\end{equation}
Specifically the oblique shear is given by $M_{ij}^{+} (i \neq j; i,j=1,2,3)$. $M_{12}^{+}$ is the oblique shear on the Galactic plane which is much larger than the oblique shears on the other two planes perpendicular to the Galactic plane. $M_{12}^{+}$ is also known as Oort Constant $B$. The details of the above equations can be found in Zhu~(\cite{Zhu00}) and Miyamoto et al~(\cite{Miyamoto93}).

We note that only $(V_{r}, V_{\ell}, V_{b})^{T}$ and $(r,\ell,b)^{T}$ measured in heliocentric Galactic coordinate system are available, $\delta \mathbf{V} = \mathbf{V} - \mathbf{V_{\bigodot}}$ and $(\delta R, \delta \theta, \delta z)^{T}$ should be transformed from $(V_{r}, V_{\ell}, V_{b})^{T}$ and $(r,\ell,b)^{T}$. Then the 12 kinematical parameters can be solved with Eq.~(\ref{eq:tylor}). Because $ \Omega_{1} $, $ \Omega_{2} $, $ M_{32}^{+} $, $ M_{33}^{+} $, $M_{22}^{+}$ $ M_{13}^{+}$, and $M_{11}^{+}$ are almost zero for disk Galaxy (Zhu~\cite{Zhu00}, Miyamoto et al.~\cite{Miyamoto93}), only $S_{1},~ S_{2}, ~S_{3}$ and Oort constants $A$, $B$ have significant values. This means that the stars in the Solar neighborhood mainly move in the direction of Galactic rotation on the Galactic plane and the Galaxy does not expand notably in radial direction. To restrict our model to these 5 main parameters, we use the 2-demensional Lindblad-Oort Model (on the Galactic plane):
\begin{eqnarray}
\kappa \mu _{\ell}\cos b & = &S_{1}\sin \ell/r-S_{2}\cos \ell/r +\Omega _{3}\cos b +M_{12}^{+}\cos b\cos 2\ell
\label{eq:SO-Mex_l}, \\
V_{r}/r & = & -S_{1}\cos \ell\cos b/r-S_{2}\sin \ell\cos b/r-S_{3}\sin b/r +M_{12}^{+}\cos ^{2}b\sin 2\ell
\label{eq:SO-Mex_r},
\end{eqnarray}
where $\kappa = 4.74047$ if the proper motion is given in the unit $\rm{mas~yr}^{-1}$, the distance in kpc and the velocity in $\rm{km~s^{-1}}$. $\mu_{\ell}$ is the proper motion in Galactic longitude. Based on the simplified Lindblad-Oort model, we obtain the Solar motion and Oort constants by a least-squares fit.

\section{Kinematical features depend on age and metallicity}
\label{sec:features}

\subsection{Selection of samples from ACAD}
\begin{figure}
\centering
\includegraphics[width=\textwidth, angle=0]{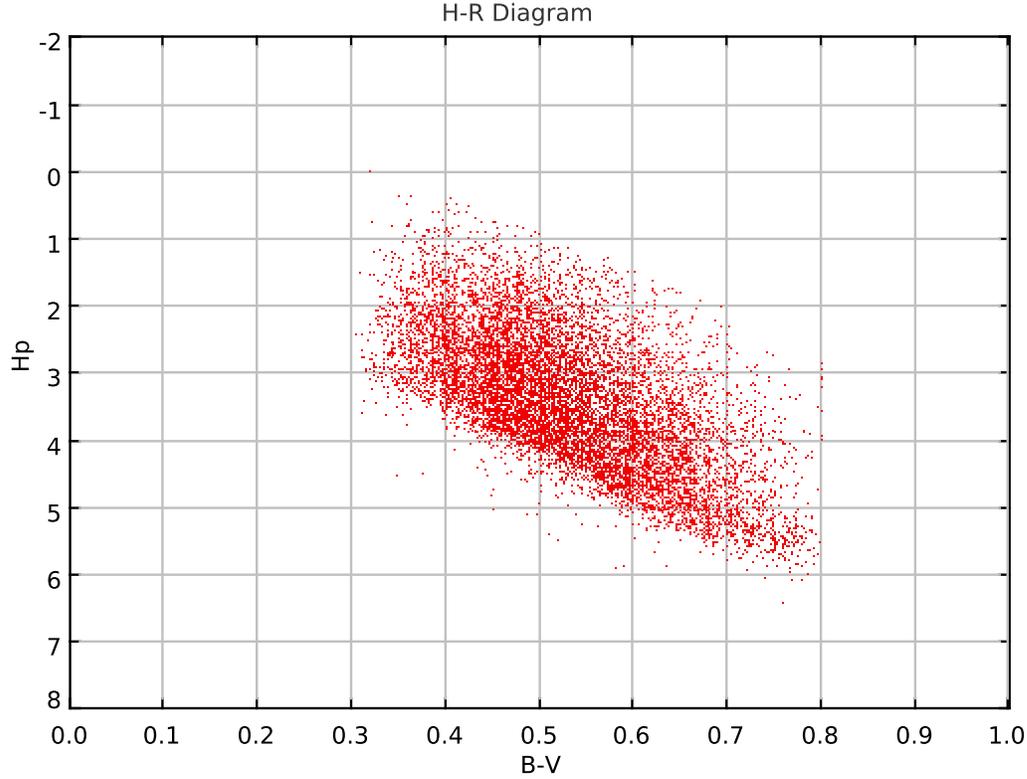}
\caption{The selected sample stars in Hertzsprung-Russel diagram, the color index $B-V$ are restricted in [0.3,0.8].}
\label{pic:H-R}
\end{figure}

The astrometric parameters (positions, distance, and proper motions) and radial velocities are tied to the Hipparcos system and IAU standard, respectively, therefore the parameters of stellar motion are homogeneous in ACAD. On the other hand, we should select stars with the distance and radial velocity data which are accurate enough. Hence, the stars revised with Hipparcos parallaxes are selected. Another important principle in selecting sample stars is that they must be main-sequence stars so that they can meet the one-to-one relation between color and absolute luminosity. We applied a similar method which was used by Walter \& Binney (\cite{Binney98}) to select the main-sequence stars with color range $0.3 < B - V < 0.8$. The selected stars on H-R diagram are plotted in Figure~\ref{pic:H-R}. Selecting the stars in the thin disk is the third principle. Since the Sun is in the thin disk and the kinematical properties of the thin disk and the thick disk are different, that of the solar neighborhood would be estimated more accurately with LSR formed by thin disk stars. Sch\"{o}nrich \& Binney~(\cite{SB09b}) found that most of the thin disk stars are younger than 6.5 Gyr. We note that number of young stars is not enough for kinematical analysis, therefore the upper limit of ages was set to be 6.5 Gyr and the lower limit was 1.5 Gyr. After these three steps of selecting, 7,810 thin disk main-sequence stars are retained for the following analysis.

\subsection{Kinematical features of the sample stars}

\subsubsection{The evolution of Solar motion and Oort Constants $A$ and $B$ with age}
In order to analyze the evolution of the kinematics, the sample stars are divided into several subsamples by age. We note that too few divisions cannot show a clear trend between kinematical parameters and age, while too many divisions would cause insufficient data in each subsample. After a few tests and comparisons, a compromised scheme was used. We divided the sample stars between 1.5 and 6.5 Gyr into five groups with a bin of 1 Gyr. Table~\ref{tbl:kinematics_age} list the variations of kinematical parameters in different age bins obtained from the least-squares fit to Eq.(\ref{eq:SO-Mex_l}) and Eq.(\ref{eq:SO-Mex_r}), also listed are the values derived from K-M giants and F3-type stars (Miyamoto \& S\^{o}ma~\cite{Miyamoto93}).

\begin{table}[width = \textwidth]
\begin{center}
\caption{Kinematical parameters in different age bins for selected F-type and G-type stars in the ACAD catalog. The parameters derived from the K-M giants and F3-type stars are also listed for comparison.}
\label{tbl:kinematics_age}
 \begin{tabular}{clrrrr}
  \hline\noalign{\smallskip}
 & unit & Min in all age bins & Max in all age bins & K-M giants & F3 stars\\
  \hline\noalign{\smallskip}
$S_1$  &  & $4.2\pm 1.2$ & $7.7\pm 1.5$ & $13.6\pm 0.3$ & $11.4\pm 0.7$ \\
$S_2$ & $\mathrm{km\ s^{-1}}$   & $11.6\pm 0.5$ & $16.7\pm 1.8$ & $23.2\pm 0.3$ & $14.1\pm 0.7$ \\
$S_3$  &  & $6.9\pm 0.8$ & $10.9\pm 1.3$ & $11.9\pm 0.3$ & $8.7\pm 0.2$\\
\hline
$A$ & $\mathrm{km~s^{-1}~kpc^{-1}}$   & $8.4\pm 0.6$ & $14.3\pm 0.9$ & $12.0\pm 0.6$ &
$13.8\pm 3.5$ \\
$B$ &  & $-20.7\pm 0.8$ & $-7.8\pm 1.9$ & $-8.6\pm 0.5$ & $-12.7\pm 2.6$ \\
  \noalign{\smallskip}\hline
\end{tabular}
\end{center}
\tablecomments{0.86\textwidth}{The K-M giants distribute from 0.5kpc to 1kpc. F3-type stars distribute from 0.1kpc to 1.3kpc. (Miyamoto \& S\^{o}ma~\cite{Miyamoto93})}
\end{table}

\begin{figure}
    \centering
    \includegraphics[width=\textwidth, angle=0]{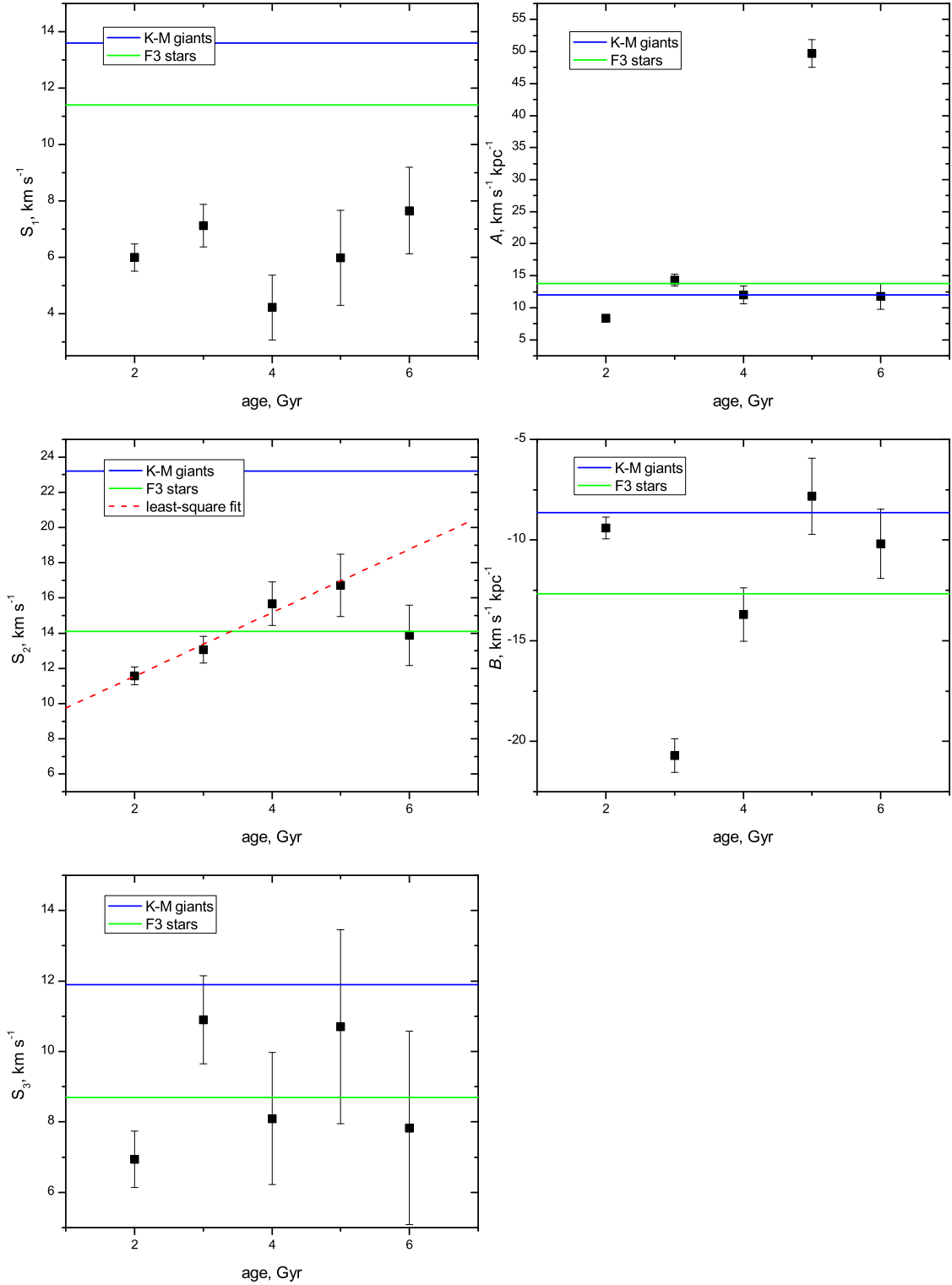}
    \caption{The age-dependence of the kinematical parameters derived from the Solar neighborhood stars. The left three panels are for the Solar motion $S_1$, $S_2$ and $S_3$. The right panels are for Oort constants $A$ and $B$. The blue line is the value derived from K-M giants and the green line is the value derived from F3 stars (Miyamoto \& S\^{o}ma~\cite{Miyamoto93}), and red dash line is the linear fit of $S_2$ against age.}
    \label{pic:Oort}
\end{figure}

In Figure~\ref{pic:Oort}, we plot the Solar motion and Oort constants derived from the stars in each age bin. $S_1$, $S_3$, Oort constant $A$ and $B$ have no systematical variation with respect to age. However we found an obvious linear increase of $S_2$ with age by $1.8\ \mathrm{km\ s^{-1}\ Gyr^{-1}}$. This can be explained by the theory that scattering processes cause the random velocities of stars to increase steadily with age (Jenkins~\cite{Jenkins92}). We also obtained $S_2 = 8.0\pm1.2\ \mathrm{km\ s^{-1}}$ by extrapolating the relation back to age = 0 Gyr.

With the relation between velocity dispersion and Solar motion, Dehnen \& Binney (\cite{Binney98}) found $S_2 = 5.3~\mathrm{km\ s^{-1}}$ for a dynamical cold disk. However Binney~(\cite{B09}) stated that the observed velocity distribution would be better fitted to the standard distribution function if the velocity dispersion could be shifted a little higher. This phenomenon implied that the $S_2$ might be underestimated. By calculating the amount ($\delta S_2 = 5.8~ \mathrm{km\ s^{-1}}$) which should be added to $S_2$, more realistic $S_2$ should be $11.0~\mathrm{km\ s^{-1}}$.

The $S_2$ parameter for the dynamical cold disk in the former studies (e.g. Dehnen \& Binney \cite{Binney98}) are based on the linear correlation between the squared velocity dispersion and Solar motion. However, this linear correlation was based on the assumption that Galactic potential is axisymmetric. In reality, the amplitude of the non-axisymmetric potential effect on $S_2$ is about $7\ \mathrm{km\ s^{-1}}$. In our study, we fitted the correlation between Solar motion and ages more directly, which shows the evolution of the Solar motion with time.

Our result is larger than the result of Dehnen \& Binney (\cite{Binney98}) and the difference is smaller than the amplitude of the non-axisymmetric component. It proves that $S_2 = 5.3\ \mathrm{km\ s^{-1}}$ is an underestimation by (Dehnen \& Binney \cite{Binney98}). At the same time, our result is smaller than that of Aumer \& Binney~(\cite{Binney09}). The possible reason is the effect of spiral arm structure (which should be reduced) in the analysis. Although our sample stars are quite close to the Sun (with 450 pc) and distribute uniformly in space, we cannot find an effective way to remove the influence of spiral structure. This effect makes the relationship between $S_2$ and age not strictly linear, so that $S_2$ for the dynamical cold disk are still underestimated. More data are needed to analyze the effect of the spiral structure on the local velocity field and to obtain the relation between Solar motion and age more precisely.

\subsection{The variation of Solar motion and Oort Constants A and B with metallicity}
The same procedure of grouping has been applied for analyzing the changes of kinematical parameters with metallicity: We divided the sample into five groups with a 0.2 [Sun] bin. As for the metallicity, since the distribution of stars centers at~0 [Sun], the very metal-poor and metal-rich populations are insufficient and would lead to a relatively large error for their kinematical parameters. It is easy to see that the errors of various parameters in metallicity bin $[-0.6,-0.4)$ are two or three times larger than that in other metallicity bins due to the insufficient stars. Therefore the results in this bin have been excluded, the remaining results are listed in Table~\ref{tbl:kinematics_metallicity}.

\begin{table}[width = \textwidth]
\begin{center}
\caption{Kinematical parameters in different metallicity bins for selected F-type and G-type stars in the ACAD catalog. The parameters derived from the K-M giants and F3-type stars are also listed for comparison.}
\label{tbl:kinematics_metallicity}
 \begin{tabular}{clrrrr}
  \hline\noalign{\smallskip}
 & unit & Min in all metallicity bins & Max in all metallicity bins & K-M giants & F3 stars\\
  \hline\noalign{\smallskip}
$S_1$  & & $5.2\pm 1.0$ & $8.6\pm 1.3$ & $13.6\pm 0.3$ & $11.4\pm 0.7$ \\
$S_2$ & $\mathrm{km\ s^{-1}}$  & $10.8\pm 1.1$ & $19.7\pm 1.3$ & $23.2\pm 0.3$ & $14.1\pm 0.7$ \\
$S_3$  &  & $6.2\pm 0.8$ & $9.5\pm 2.1$ & $11.9\pm 0.3$ & $8.7\pm 0.2$\\
\hline
$A$ & $\mathrm{km~s^{-1}~kpc^{-1}}$  & $2.3\pm 1.2$ & $17.7\pm 0.7$ & $12.0\pm 0.6$ &
$13.8\pm 3.5$ \\
$B$ &  & $-19.7\pm 0.6$ & $-4.0\pm 1.4$ & $-8.6\pm 0.5$ & $-12.7\pm 2.6$ \\
  \noalign{\smallskip}\hline
\end{tabular}
\end{center}
\tablecomments{0.86\textwidth}{The K-M giants distribute from 0.5kpc to 1kpc. F3-type stars distribute from 0.1kpc to 1.3kpc. (Miyamoto \& S\^{o}ma~\cite{Miyamoto93})}
\end{table}

\begin{figure}
   \centering
   \includegraphics[width=\textwidth, angle=0]{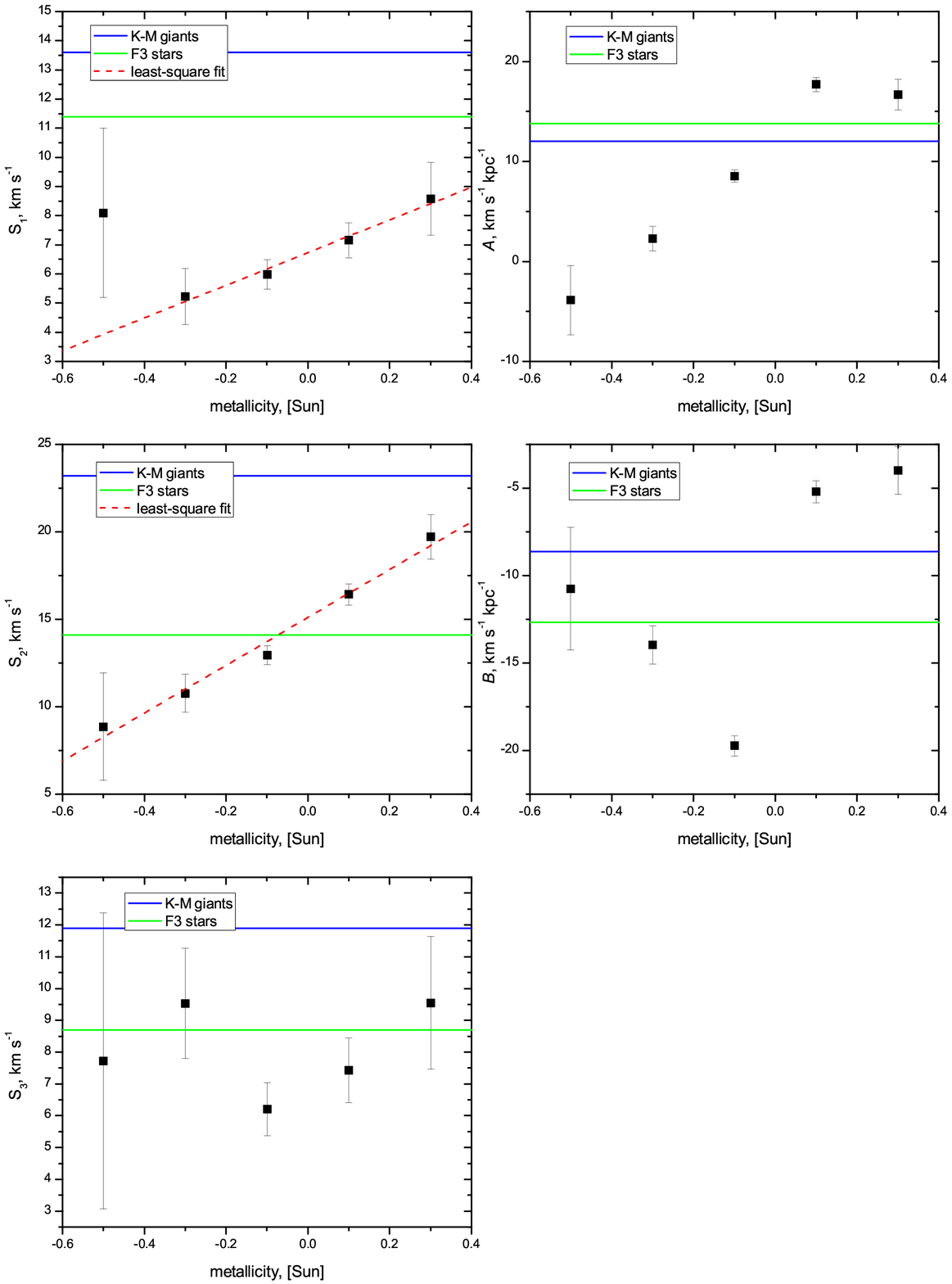}
   \caption{The matellicity-dependency of the kinematical parameters derived from the Solar neighborhood stars. The left three panels are for the Solar motion$S_1$, $S_2$ and $S_3$. The right two panels are for Oort constants $A$ and $B$. The blue line is the value derived from K-M giants and the green line is the value derived from F3 stars (Miyamoto \& S\^{o}ma~\cite{Miyamoto93}). The red dash line is the linear fit against metallicity.}
   \label{pic:OortMetal}
\end{figure}

We plot the variations of the kinematical parameters against metallicity in Figure~\ref{pic:OortMetal}. A similar linear relation can be found between $S_2$ and metallicity. There is also a clear linear relation between $S_1$ and metallicity by ignoring the value in metallicity bin $[-0.6,-0.4)$ which has a large fitting error. The rates of $S_{1}$ and $S_{2}$ are $0.6\ \mathrm{km\ s^{-1}}/\ 0.1\mathrm{[Sun]}$ and $1.4\ \mathrm{km\ s^{-1}}/\ 0.1\mathrm{[Sun]}$, respectively. By interpolating, we found that $S_{1} = 6.7\ \mathrm{km\ s^{-1}}$ and $S_{2} = 15.1\ \mathrm{km\ s^{-1}}$ at [Fe/H] = 0 [Sun] (metallicity of the Sun).

\begin{figure}
    \centering
    \includegraphics[width=\textwidth, angle=0]{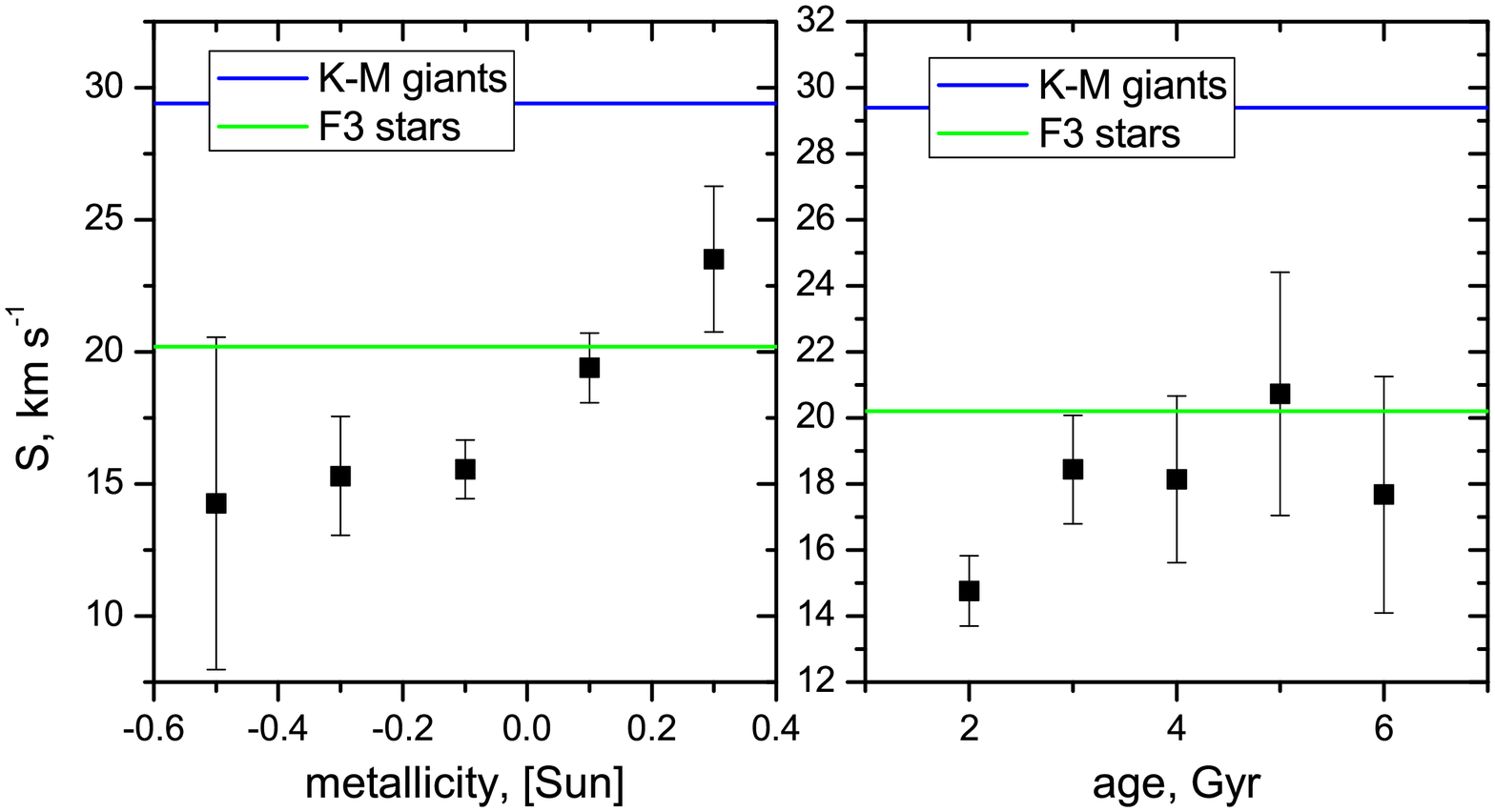}
    \caption{The total Solar motion $S_{\bigodot}$ versus age (right) and metallicity (left). The blue line is the $S_{\bigodot}$ derived from K-M giants and green line is the $S_{\bigodot}$ derived from F3-type stars (Miyamoto \& S\^{o}ma~\cite{Miyamoto93}).}
    \label{pic:STotal}
\end{figure}

Sch\"{o}nrich et al.~(\cite{SB09a}) found that stellar migration is the main mechanism for the metallicity gradient in the Solar neighborhood. Our results for $S_1$ and $S_2$ agree well with the conclusion that radial stellar migration causes both non-circular orbits and guiding-centre shifts (Sellwood \& Binney~\cite{Sellwood02}). Since the total Solar motion ($S_{\bigodot}$) can reflect the status of asymmetric drifts, our results also proved that asymmetric drifts are stronger for the metal-rich populations in Figure~\ref{pic:STotal} (Sch\"{o}nrich et al.~\cite{Schonrich10}). In the figure, $S_{\bigodot}$ increase from $14.3\pm 6.3\ \mathrm{km\ s^{-1}}$ to $23.5\pm 2.8\ \mathrm{km\ s^{-1}}$ with respect to metallicity. Nevertheless, $S_{\bigodot}$  does not show a clear systematical variation against age.

\subsubsection{Rotation of the Sun around the Galactic center}
The angular velocity of the Sun around the Galactic center can be described by the difference between the Oort constants $A$ and $B$:
\begin{equation}
A - B = M_{12}^{+} - \Omega_3 = \dfrac{V_{\theta}}{R} - \dfrac{1}{R}\dfrac{\partial V_{R}}{\partial \theta}.
\label{eq:A-B}
\end{equation}

\begin{figure}
    \centering
    \includegraphics[width=\textwidth, angle=0]{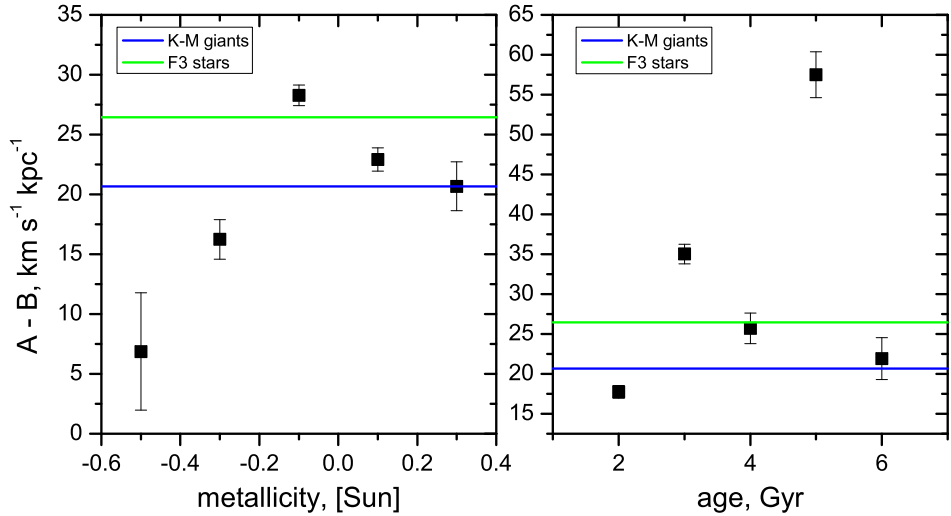}
    \caption{$A - B$ versus age (right) and metallicity (left). The blue line is $A - B$ derived from K-M giants and green line is the $A - B$ derived from F3-type stars (Miyamoto \& S\^{o}ma~\cite{Miyamoto93}).}
    \label{pic:A_B}
\end{figure}

If we assume the motions of the stars in the Solar neighborhood is circular, we can obtain $\partial V_{R}/\partial \theta \simeq 0$ and $A - B = V_{\theta}/R$. Figure~\ref{pic:A_B} shows the variation of $A-B$ with respective to the metallicity (left panel) and age (right panel), from which no obvious trend can be found. The medium values of $A-B$ are $20.7\pm2.0$ and $25.7\pm1.9~{\rm km~s^{-1}~kpc^{-1}}$ for age and metallicity grouping, respectively. Adopting $R_0=8.5 {\rm kpc}$ as the distance from the Sun to the Galactic center, we calculated the rotational velocity of the Sun around the Galactic center with $A - B$ derived from different groups of stars, which gives $V_0=175.9\pm16.4~{\rm km~s^{-1}}$ and $V_0=218.5\pm17.3~{\rm km~s^{-1}}$ for for age and metallicity grouping. The discrepancy in the these results between the two kinds of groupings is due to the fact that the age and metallicity are not strictly correlated.

\section{Discussion and Conclusion}
\label{sec:discussion}
In this paper we collected three best available catalogs of F-type and G-type stars with astrophysical data. After revising the distance, supplying radial velocity and photometric data, we built The Astrometric catalog associated with astrophysical data (ACAD). It is a comprehensive catalog using 8 different catalogs. The total number of stars in ACAD is 27,553. ACAD can be useful for the further research on the Galactic kinematics. It can also serve as an adjustment for the future astrophysical data. The complete ACAD is available as an electronic table. Detailed descriptions can be found in Appendix~\ref{sec: appendix} and Table~\ref{tbl:catalog}.

To analyze the kinematics of the stars on the Galactic disk, we used the two-dimensional Lindblad-Oort Model to estimate the three dimensional Solar motion and Oort constants $A$ and $B$. In our study, axisymmetric Galactic potential is no longer a pre-condition and we found that $S_2$ increase linearly with age which may be a result of scattering process. By extrapolating this relationship back to 0 Gyr, $S_2$ is found to be $8.0\pm1.2~\mathrm{km~s^{-1}}$ for a dynamical cold disk. This value is in the middle of the value ($S_2=5.3\ \mathrm{km\ s^{-1}}$) calculated by Dehnen \& Binney(\cite{Binney98}) and the value ($S_2=11.0\ \mathrm{km\ s^{-1}}$) calculated by Aumer \& Binney(\cite{Binney09}). It shows that spiral structure still affect the estimation of $S_2$.

We also found an increase of $S_1$ and $S_2$ with respect to metallicity, which indicates that radial stellar migration leads to both non-circular orbits and guiding-centre shifts (Sellwood \& Binney~\cite{Sellwood02}). This effect is more significant along the rotating direction ($\mathrm{d}S_2/{\rm d}\mathrm{[Fe/H]}=1.37\ \mathrm{km\ s^{-1}}/0.1\mathrm{[Sun]}$) than along the radius direction ($\mathrm{d}S_1/{\rm d}\mathrm{[Fe/H]}=0.56\ \mathrm{km\ s^{-1}}/0.1\mathrm{[Sun]}$). By interpolating, we obtained $S_{1} = 6.7\ \mathrm{km\ s^{-1}}$ and $S_{2} = 15.1\ \mathrm{km\ s^{-1}}$ for the stars at Solar metallicity.

By using Oort Constants $A$ and $B$, the rotational speed of the Sun around the Galactic center was calculated with different groups of stars. It shows that the Local Standard of Rest (LSR) which is formed by stars of different ages or metallicities does not affect systematically the calculation of the rotational velocity of the Sun.

Before the data release from the next astrometric satellite Gaia, which is expected to provide billions of stars with extremely high accurate astrometric parameters, radial velocities and photometric data, there is still a gap of more than ten years for us to carry out prospective studies and prepare for future applications of Gaia data. We will update ACAD after collecting more radial velocity, age and metallicity data in the future.

\begin{acknowledgements}
This work was funded by the National Natural Science Foundation of China (NSFC) under grant No.11173014. and by the Natural Science Foundation of Jiangsu Province under No. SBK201341200.
\end{acknowledgements}

\appendix                  

\section{Data description of The Astrometric catalog associated with Astrophysical Data (ACAD)}
\label{sec: appendix}
The detailed data arrangement of the compiled catalog ACAD is given in Table~\ref{tbl:catalog}. The epoch of the catalog is J2000.0 (ICRS). Explanation for several indexes in ACAD:
(1) FD [F,P,T] represents the distances calculated from different methods:\\
\begin{description}
\item[F :] Fitting distance calculated from proper motion and color.
\item[T :] Parallax distance calculated from trigonometric parallax.
\item[P :] Photometric distance calculated from absolute magnitude and visual magnitude.
\end{description}
(2) FA [1,3] represents the different sources of astrophysical data:
\begin{enumerate}
\item Geneva-Copenhagen survey of the Solar neighborhood (GCS, Nordstr\"{o}m et al.~\cite{GCS}).
\item Catalog of ages, metallicities, orbital elements and other parameters for nearby F stars (MS95, Marsakov and Shevelev~\cite{v89}).
\item The N2K Consortium. IV. New temperatures and metallicities for more than 100,000 FGK dwarfs (AM06, Ammons S. M. et al.~\cite{v136}).
\end{enumerate}
(3) FV[1-6] (flag of space velocity) represents the different combination of proper motion and redial velocity, all the possibility are listed in Table~\ref{tbl:RVstatistic}.

\newpage

\begin{table}
\begin{center}
\caption[]{ Contents of The Astrometric catalog associated with astrophysical data}
\label{tbl:catalog}
\begin{tabular}{llll}
  \hline\noalign{\smallskip}
No &  Label      & Unit & Description                    \\
  \hline\noalign{\smallskip}
 1 & GLON        & Deg    & Galactic longitude \\
 2 & GLAT       & Deg    & Galactic latitude \\
 3 & MV       & mag    & Absolute magnitude \\
 4 & DIST       & pc    & Heliocentric distance of the star \\
 5 & UDIST         & pc    & ?=99999.999 Upper confidence limit on heliocentric distance \\
 6 & LDIST         & pc    & ?=99999.999 Lower confidence limit on heliocentric distance \\
 7 & FD        & ---    &  [F,P,T] Flag of distance \\
 8 & AGE       & Gyr & ?=99.99 Age \\
 9 & UAGE       & Gyr & ?=99.99 Upper confidence limit on age \\
10 & LAGE      & Gyr    & ?=99.99 Lower confidence limit on age \\
11 & FE/H      & [Sun]    & Metallicity \\
12 & UFE/H     & [Sun]    & ?=999.999 Upper confidence limit on metallicity \\
13 & LFE/H    & [Sun] & ?=999.999 Lower confidence limit on metallicity \\
14 & FA    & --- & [1,3] Flag of astrophysical data \\
15 & U         & km/s    & Heliocentric space velocity component in the radial direction\\
16 & V      & km/s    & Heliocentric space velocity component in the direction of rotation \\
17 & W         & km/s    & Heliocentric space velocity component perpendicular to the Galactic plane \\
18 & FV      & ---    & [1,6] Flag of space velocity \\
19 & SP         & ---    & ?=--- Spectral type in MK classification \\
  \noalign{\smallskip}\hline
\end{tabular}
\end{center}
\end{table}

\label{lastpage}

\end{document}